\let\includefigures=\iftrue
%
\let\useblackboard=\iftrue
%
%
\newfam\black
\input harvmac
\noblackbox
\includefigures
\message{If you do not have epsf.tex (to include figures),}
\message{change the option at the top of the tex file.}
\input epsf
\def\figin{\epsfcheck\figin}\def\figins{\epsfcheck\figins}
\def\epsfcheck{\ifx\epsfbox\UnDeFiNeD
\message{(NO epsf.tex, FIGURES WILL BE IGNORED)}
\gdef\figin##1{\vskip2in}\gdef\figins##1{\hskip.5in}
\else\message{(FIGURES WILL BE INCLUDED)}%
\gdef\figin##1{##1}\gdef\figins##1{##1}\fi}
\def\DefWarn#1{}
\def\figinsert{\goodbreak\midinsert}
\def\ifig#1#2#3{\DefWarn#1\xdef#1{fig.~\the\figno}
\writedef{#1\leftbracket fig.\noexpand~\the\figno}%
\figinsert\figin{\centerline{#3}}\medskip\centerline{\vbox{
\baselineskip12pt\advance\hsize by -1truein
\noindent\footnotefont{\bf Fig.~\the\figno:} #2}}
\bigskip\endinsert\global\advance\figno by1}
\else
\def\ifig#1#2#3{\xdef#1{fig.~\the\figno}
\writedef{#1\leftbracket fig.\noexpand~\the\figno}%
\global\advance\figno by1}
\fi
%

\def\smallfig#1#2#3{\DefWarn#1\xdef#1{fig.~\the\figno}
\writedef{#1\leftbracket fig.\noexpand~\the\figno}%
\figinsert\figin{\centerline{#3}}\medskip\centerline{\vbox{
\baselineskip12pt\advance\hsize by -1truein
\noindent\footnotefont{\bf Fig.~\the\figno:} #2}}
\endinsert\global\advance\figno by1}

\useblackboard
\message{If you do not have msbm (blackboard bold) fonts,}
\message{change the option at the top of the tex file.}
\font\blackboard=msbm10 scaled \magstep1
\font\blackboards=msbm7
\font\blackboardss=msbm5
\textfont\black=\blackboard
\scriptfont\black=\blackboards
\scriptscriptfont\black=\blackboardss

\else

\fi
%
\def\yboxit#1#2{\vbox{\hrule height #1 \hbox{\vrule width #1
\vbox{#2}\vrule width #1 }\hrule height #1 }}
\def\fillbox#1{\hbox to #1{\vbox to #1{\vfil}\hfil}}
\def\ybox{{\lower 1.3pt \yboxit{0.4pt}{\fillbox{8pt}}\hskip-0.2pt}}
%
%

\def\comments#1{}

\def\Re{{\rm Re\hskip0.1em}}

\def\CC{{\cal C}}

\def\CN{{\cal N}}
\def\CO{{\cal O}}

\def\CS{{\cal S}}


\def\II{\relax{I\kern-.10em I}}

\def\IZ{\relax\ifmmode\mathchoice
{\hbox{\cmss Z\kern-.4em Z}}{\hbox{\cmss Z\kern-.4em Z}}
{\lower.9pt\hbox{\cmsss Z\kern-.4em Z}}
{\lower1.2pt\hbox{\cmsss Z\kern-.4em Z}}
\else{\cmss Z\kern-.4emZ}\fi}
\def\IB{\relax{\rm I\kern-.18em B}}
\def\IC{{\relax\hbox{$\inbar\kern-.3em{\rm C}$}}}
\def\ID{\relax{\rm I\kern-.18em D}}
\def\IE{\relax{\rm I\kern-.18em E}}
\def\IF{\relax{\rm I\kern-.18em F}}
\def\IG{\relax\hbox{$\inbar\kern-.3em{\rm G}$}}
\def\IGa{\relax\hbox{${\rm I}\kern-.18em\Gamma$}}
\def\IH{\relax{\rm I\kern-.18em H}}
\def\II{\relax{\rm I\kern-.18em I}}
\def\IK{\relax{\rm I\kern-.18em K}}
\def\IP{\relax{\rm I\kern-.18em P}}

%

\def\inbar{\,\vrule height1.5ex width.4pt depth0pt}

\font\cmss=cmss10 
\def\IR{\relax{\rm I\kern-.18em R}}

\def\into{\hookrightarrow}
%


%

\def\lp10{\ell_p^{10}}
\def\lp11{\ell_p^{11}}
\def\R11{R_{11}}

\def\frac#1#2{{#1 \over #2}}


\def\cf{{\it c.f.}}
\def\etal{{\it et. al.}}
\hyphenation{Di-men-sion-al}

\def\np{{\it Nucl. Phys.}}

\lref\bdlr{I. Brunner, M.R. Douglas, A. Lawrence and
C. R\"omelsberger, ``D-branes on the Quintic,'' hep-th/9906200.}
\lref\bbs{K.~Becker, M.~Becker and A.~Strominger,
``Five-branes, membranes and nonperturbative string theory,"
Nucl. Phys. {\bf B456} (1995) 130, hep-th/9507158.}
\lref\bsv{M. Bershadsky, V. Sadov and C. Vafa, ``D-branes and
Topological Field Theories'', \np\ {\bf B463}\ (1996) 420, 
hep-th/9511222.}
\lref\thomas{R.P. Thomas, ``An obstructed bundle on a Calabi-Yau 3-fold,''
math.AG/9903034.}
\lref\ooy{H. Ooguri, Y. Oz and Z. Yin, ``D-branes
on Calabi-Yau spaces and their mirrors,''
Nucl. Phys. {\bf B477} (1996) 407, hep-th/9606112.}
\lref\diacrom{D.-E. Diaconsescu and C. R\"omelsberger,
``D-branes and bundles on elliptic fibrations'',
\np\ {\bf B574} (2000) 245; hep-th/9910172.}
\lref\shamitjohn{S. Kachru and J. McGreevy, ``Supersymmetric
three-cycles and supersymmetry breaking,'' Phys. Rev. {\bf D61} (2000)
026001, hep-th/9908135.}

\lref\mikerev{M. R. Douglas, ``Topics in D Geometry,'' 
Class. Quant. Grav. {\bf 17} (2000) 1057, hep-th/9910170.} 
\lref\noncomp{M. Douglas, B. Fiol and C. Romelsberger, ``The Spectrum of BPS
Branes on a Noncompact Calabi-Yau Space,'' hep-th/0003263.}
\lref\govind{S. Govindarajan and T. Jayaraman, ``On the Landau-Ginzburg
Description of Boundary CFTs and Special Lagrangian Submanifolds,'' 
hep-th/0003242.}
\lref\brunner{I. Brunner and V. Schomerus, ``D-branes at Singular Curves
of Calabi-Yau Compactifications,''JHEP {\bf 0004} (2000) 020, 
hep-th/0001132.} 
\lref\greene{B. Greene and C. Lazaroiu, ``Collapsing D-branes in 
Calabi-Yau Moduli Space I,'' hep-th/0001025.} 
\lref\naka{M. Naka, M. Nozaki,
``Boundary states in Gepner models,''
JHEP \ {\bf 0005} (2000) 027, hep-th/0001037.}
\lref\lazar{C. Lazaroiu, ``Collapsing D-branes in One Parameter Models
and Small/Large Radius Duality,'' hep-th/0002004.}
\lref\denef{F. Denef, ``Supergravity Flows and D-brane Stability,''
hep-th/0005049.} 
\lref\horietal{K. Hori, A. Iqbal and C. Vafa, ``D-branes and
Mirror Symmetry,'' hep-th/0005247.}
\lref\scheidd{E. Scheidegger, ``D-branes on Some One Parameter 
and Two Parameter
Calabi-Yau Hypersurfaces,''  JHEP \ {\bf 0004} (2000) 003, 
hep-th/9912188.}
\lref\lerche{P. Kaste, W. Lerche and C. Lutken, 
``D-branes on K3 Fibrations,''
hep-th/9912147.} 
\lref\kklm{S. Kachru, S. Katz, 
A. Lawrence and J. McGreevy, ``Open String Instantons
and Superpotentials,'' hep-th/9912151.}
\lref\stable{M. Douglas, B. Fiol and C. Romelsberger, ``Stability and BPS
Branes,'' hep-th/0002037.}

\lref\syz{A. Strominger, S.-T. Yau and E. Zaslow, ``Mirror
Symmetry is T-Duality,'' Nucl. Phys. {\bf B479}\ (1996) 243, hep-th/9606040.}
\lref\konthom{M. Kontsevich, ``Homological algebra of
mirror symmetry,'' Proc. of the 1994 International
Congress of Mathematicians, Birkh\"auser (Boston) 1995, 
alg-geom/9411018.}
\lref\bundlemir{C. Vafa, ``Extending mirror conjecture
to Calabi-Yau with bundles,'' hep-th/9804131.}
\lref\kmp{S. Katz, D.R. Morrison and M.R. Plesser,
``Enhanced gauge symmetry in type II string theory,''
Nucl. Phys. {\bf B477} (1996) 105, hep-th/9601108.}
\lref\twopI{P. Candelas, X. de la Ossa, A. Font,
S. Katz and D.R. Morrison, ``Mirror symmetry
for two-parameter models -- I,'' Nucl. Phys. {\bf B416} (1994)
481, hep-th/9308083.}
\lref\gp{B. Greene and M. Plesser, ``Duality in Calabi-Yau Moduli Space,''
Nucl. Phys. {\bf B338} (1990) 15.}
\lref\wittmirr{E. Witten, ``Mirror manifolds and
topological field theory,'' in {\it Mirror Symmetry I},
S.-T. Yau (ed.), American Mathematical Society (1998), 
hep-th/9112056.}
\lref\edphases{E. Witten, ``Phases of N=2 theories in two dimensions,'' 
Nucl. Phys. {\bf B403} (1993) 159, hep-th/9301042.}
\lref\mdmm{P.S. Aspinwall, B.R. Greene and D.R. Morrison,
``The monomial-divisor mirror map,'' Internat. Math.
Res. Notices {\bf 93} (1993) 319, alg-geom/9309007.}
\lref\konthom{M. Kontsevich, ``Homological algebra of
mirror symmetry,'' Proc. of the 1994 International
Congress of Mathematicians, Birkh\"auser (Boston) 1995, 
alg-geom/9411018.}

\lref\candetal{P. Candelas, X. de la 
Ossa, P. Green and L. Parkes, ``A Pair of Calabi-Yau
Manifolds as an Exactly Soluble Superconformal 
Theory,'' Nucl. Phys. {\bf B359} (1991) 21.}

\lref\ov{H. Ooguri and C. Vafa, ``Knot
invariants and topological strings,'' hep-th/9912123.}

\lref\mclean{R.~McLean, ``Deformations of Calibrated Submanifolds'',
Duke Univ. PhD thesis,
Duke preprint 96-01: see www.math.duke.edu/preprints/1996.html.}
\lref\hitchin{N.J.~Hitchin, ``The moduli space of special Lagrangian
submanifolds,''  Ann. Scuola Norm. Sup. Pisa Cl. Sci. (4) {\bf 25}
(1997) 503, dg-ga/9711002.}
\lref\joyceslag{D. Joyce, ``On counting special Lagrangian
homology 3-spheres,'' hep-th/9907013.}

\lref\katzone{S. Katz, ``On the finiteness
of rational curves on quintic threefolds,''
Comp. Math. {\bf 60}, 151 (1986).}
\lref\katztwo{S. Katz, ``Rational Curves
on Calabi-Yau Threefolds,'' in {\it Mirror
Symmetry I} (S.-T. Yau, ed.), American
Mathematical Society and International Press (1999), 
alg-geom/9312009.}
\lref\reid{M. Reid, ``Minimal Models
of Canonical 3-folds'', pp. 131-180,
Advanced Studies in Pure Mathematics 1, ed. S. Iitaka,
Kinokuniya (1983).}
\lref\burns{D. Burns, ``Some background and examples
in deformation theory,'' in {\it Complex Manifold Techniques in
Theoretical Physics}, D. Lerner and P. Sommers Eds., Pitman (1979).}
\lref\friedman{R. Friedman, ``Simultaneous resolution of threefold 
double points'', Math. Ann. 274 (1986), 671--689.}
\lref\bkl{J. Bryan, S. Katz, and N.C. Leung, ``Multiple covers and 
the integrality conjecture for rational curves in Calabi-Yau threefolds,''
math.AG/9911056.}
\lref\kodaira{K. Kodaira, ``A Theorem of Completeness of
Characteristic Systems for Analytic Families of Compact
Submanifolds of Compact Manifolds,'' Ann. Math. {\bf 75} (1962) 146.}
\lref\yau{S.-T. Yau, ``Calabi's conjecture and some
new results in algebraic geometry,'' Proc. Nat. Acad. Sci.
U.S.A. {\bf 74} (1977) 1798.}

\lref\wittcs{E. Witten, ``Chern-Simons gauge theory
as a string theory,'' in {\it The Floer Memorial
Volume}, H. Hofer \etal, eds., Birkhauser (1995), Boston, 
hep-th/9207094.}

\lref\gregjeffinst{J.A. Harvey and
G. Moore, ``Superpotentials and membrane instantons,'' 
hep-th/9907026.}

\lref\bpsalg{J.A. Harvey and G. Moore,
``On the algebras of BPS states,'' Comm. Math. Phys. {\bf 197}
(1998) 489, hep-th/9609017.}
\lref\marinoetal{M. Marino, R. Minasian, G. Moore
and A. Strominger, ``Nonlinear instantons from
supersymmetric $p$-branes,'' JHEP {\bf 0001} (2000) 005, 
hep-th/9911206.}

\lref\braneprobe{M.R. Douglas and M. Li, ``D-brane realization
of $\CN=2$ super Yang-Mills theory in four
dimensions,'' hep-th/9604041; A. Sen,
``F-theory and orientifolds,'' Nucl. Phys. {\bf B475} (1996) 562,
hep-th/9605150; T. Banks, M.R. Douglas and
N. Seiberg, ``Probing F-theory with branes,'' Phys. Lett. {\bf B387}
(1996) 278, hep-th/9605199; N. Seiberg, ``IR dynamics on
branes and space-time geometry,'' Phys. Lett. {\bf B384} (1996) 81,
hep-th/9606017.}

\lref\bdfm{T. Banks, L.J. Dixon, D. Friedan and
E. Martinec, ``Phenomenology and conformal field theory:
or, can string theory predict the weak mixing angle?,''
Nucl. Phys. {\bf B299} (1988) 613.}

\Title{\vbox{\baselineskip12pt\hbox{hep-th/0006047}
\hbox{SU-ITP-00/15}
\hbox{SLAC-PUB-8461}
\hbox{OSU-M-99-7}}}
{\vbox{
\centerline{Mirror symmetry for open strings}}}
\smallskip
\centerline{Shamit Kachru$^{1,2}$, Sheldon Katz$^{3}$,}
\medskip
\centerline{Albion Lawrence$^{1,2}$  and John McGreevy$^{1}$}
\bigskip
\bigskip 
\centerline{$^{1}${Department of Physics,
Stanford University, Stanford, CA 94305}}
\medskip
\centerline{$^{2}${SLAC Theory Group, MS 81, PO Box 4349, Stanford,
CA 94309}}
\medskip
\centerline{$^{3}${Department of Mathematics,
Oklahoma State University, Stillwater, OK 74078}}
\bigskip
\bigskip
\noindent

We discuss the generation of superpotentials in
$d=4$, $\CN=1$ supersymmetric field theories
arising from type IIA D6-branes wrapped on supersymmetric
three-cycles of a Calabi-Yau threefold.
In general, nontrivial 
superpotentials arise from sums over disc instantons.
We then find several examples of special Lagrangian
three-cycles with nontrivial topology which are mirror to
obstructed rational curves, conclusively 
demonstrating the existence of such instanton effects.
In addition, we present explicit examples of disc instantons
ending on the relevant three-cycles.
Finally, we give a preliminary construction of a
mirror map for the open string moduli, in a large-radius
limit of the type IIA compactification.

\Date{June 2000}

\newsec{Introduction}

{\it ``The importance of instanton computations in string theory
and in M-theory can hardly be overstated.''\gregjeffinst

\hfill -- J.A. Harvey and G. Moore}

\medskip

There are many important motivations for studying 
the physics of D-branes on Calabi-Yau
threefolds in type II string theories (or orientifolds thereof). 
To begin with, space-filling branes
provide a microscopic construction 
of brane world models with ${\cal N}=1$ supersymmetry.
In addition, physical objects in these theories
(such as the moduli space of vacua and the superpotential)
have a geometric expression.  Hence, these theories provide
a rich new context for studying quantum geometry
via $\CN=1$ field theories, along the lines of previous work
on $\CN=2$ brane probe theories \braneprobe.
For a fairly recent introductory review, see \mikerev; recent work on 
this subject has appeared e.g. in \refs{\bdlr,
\lerche,\kklm,\scheidd,\naka,\greene,
\brunner,\lazar,\stable,\govind,\noncomp,\denef,\horietal}.  

Consider a compactification of type IIA string theory on a Calabi-Yau
threefold $M$.  A single D6 brane wrapped
on a supersymmetric three-cycle
$\Sigma \subset M$ realizes a 4d ${\cal N}=1$ 
quantum field theory.\foot{To avoid RR
tadpoles, one can take $M$ to be noncompact, 
or consider a full orientifold
model which also has orientifold planes.  Alternatively, 
since we will be working at tree level, one can consider
a non-space filling brane whose worldvolume theory still has 4 supercharges,
and view the superpotentials we compute in that context.} 
In \kklm, we began to explore the consequences of mirror symmetry for such
brane worldvolume theories (related work appears in \refs{\wittcs,
\bundlemir,\ov}).  
We found that the
moduli space of vacua has complex dimension $b_1(\Sigma)$, to
all orders in $\sigma$-model perturbation theory.  Any superpotential
must be generated by nonperturbative
worldsheet effects, i.e. disc instantons.
Now, choose $\Sigma$ so that the mirror cycle $\CC$ is a rational
curve in the mirror threefold $W$.  The mirror of
the above D6-brane is a D5-brane wrapped on $\CC\times\IR^4$.
When $\CC$ has obstructed first-order deformations, the
deformation is described by a massless scalar field with a
higher-order superpotential; this superpotential is described
exactly by classical geometry 
\refs{\bdlr,\kklm}.  Mirror symmetry implies
a disc instanton-generated
superpotential for the IIA D6 brane.  Ideally we
could use this to compute the instanton sum exactly.
The first obstacle to this program is that
the explicit construction of
such D-brane mirror pairs is quite difficult, 
and examples of
compact 
special Lagrangian three-cycles with $b_{1}(\Sigma)\neq 0$
have been scarce.  

In this paper, we further this program by providing
examples of such pairs, and developing a preliminary
understanding of the structure of the instanton sums and the
mirror map.  We begin in \S2 with a more detailed review of  
supersymmetric D-branes in Calabi-Yau compactifications.
These have a standard 
classification as A-type or B-type branes \ooy,
which roughly correspond to special Lagrangian cycles
and holomorphic cycles, respectively. 
The superpotentials on B-type branes arise 
from classical geometry \refs{\bdlr,\kklm} and we 
review the geometry of a few specific examples 
(some with nontrivial superpotentials and some without).   We also
discuss the qualitative features of  
superpotentials for A type branes.  
In \S3, we give an explicit construction 
of the special Lagrangian three cycles
which are mirror to the explicit examples of B-type branes discussed in \S2.
In particular, we find examples of smooth three-cycles with
nonvanishing $b_1$ whose 
mirrors have moduli space dimension less than $b_1$.
This effectively proves the existence of disc instanton-generated
superpotentials.  We also give explicit examples
of disc instantons, i.e. holomorphic discs with boundary in a nontrivial
homology class on the special Lagrangian cycle. 
In \S4, we use mirror symmetry to make some statements
about the instanton-generated superpotential for
our A-type examples.  We first discuss 
a mechanism by which disc instanton effects in our examples could
(partially) cancel at special loci in closed string moduli space. 
We then give a preliminary description of
the mirror map for open string moduli in one
example.
We close with a discussion of promising future directions in \S5. 

\newsec{Superpotentials from D-branes} 

\subsec{A-type and B-type branes}

There are two distinct classes of supersymmetric branes in Calabi-Yau
compactifications: A-type and B-type branes \ooy\
(which can be constructed as boundary states in the topological
A- and B-twisted sigma models respectively,
following the notation of \wittmirr).
To help distinguish between these cases, we will denote by $M$ 
a Calabi-Yau used for studying A-type branes, and
by $W$ a Calabi-Yau used for studying B-type branes.  When we give examples
in later sections, mirror pairs will be 
identified by using a common subscript, 
$(M_i,W_i)$.  In geometric language, 
B-type branes correspond to branes wrapped on holomorphic
0,2,4 and 6-cycles of a Calabi-Yau $W$; while
A type branes correspond to branes wrapped on a special
Lagrangian three-cycle $\Sigma \subset M$.   
In both cases, one has to choose a gauge
field configuration on the D-brane; 
the supersymmetry-preserving bundles correspond
to flat bundles for A-type branes and to
stable, holomorphic bundles
for B-type branes.\foot{We are being schematic.
A more precise discussion of B-type branes
as coherent sheaves can be found in \bpsalg;
supersymmetric configurations with NS 2-form moduli turned on
can be found in \marinoetal.}

Assuming the branes to be space-filling,
one can prove the following general results about the dependence
of the $\CN=1$ brane worldvolume action on the Calabi-Yau moduli: 

\item{$\bullet$} 
The superpotential for B-type branes depends only on the
complex structure moduli, while FI terms depend 
only on K\"ahler moduli.

\item{$\bullet$}
The mirror story holds for A-type branes:  the superpotential
depends only on K\"ahler moduli, while the FI terms depend 
on complex structure moduli.

\noindent
The statements about the superpotential 
were proven using worldsheet 
techniques in \bdlr.  The correspondence
between FI terms and Calabi-Yau moduli has 
been explored in \refs{\mikerev,
\shamitjohn}; explicit examples of 
superpotentials for B-type branes were given in 
\refs{\bdlr,\kklm}. 

As with closed string $\sigma$-models, the 
open string $\sigma$-model coupling constants (in which one expands
$\sigma$-model perturbation theory)
are related to the choice of the K\"ahler 
form on $W$, and not to the complex
structure.  It follows that 
for B-type branes one can determine the superpotential
exactly at $\sigma$-model tree level,
using classical geometry.  In contrast, 
for A-type branes (at least for a single 
brane, which is the case of interest to
us here) the superpotential is entirely determined 
by ``stringy'' disc instanton corrections \kklm. 

\subsec{Superpotentials for B-type branes}

Information about the superpotential for $B$-type
branes is contained in the deformation theory
for these branes (and for the gauge bundles
on those branes).  We will review here the case
of branes wrapping curves in a threefold 
\refs{\bdlr\kklm}, since these are 
the examples we use in this paper.

For a holomorphic curve $\CC$ in a Calabi-Yau threefold
$W$, the number of first-order holomorphic deformations is
$d = \dim H^0(\CC, \CN_\CC)$, where $\CN_{\CC}$
is the normal bundle of $\CC\subset W$.  For
a single D5-brane wrapping $\CC$ in type IIB string theory,
this leads to $d$ massless neutral chiral supermultiplets
(in addition to the $U(1)$ vector multiplet).
A superpotential for chiral multiplets naturally
corresponds to an  obstruction to extending
the associated first-order deformations to higher order.  

Geometrically, the obstruction can only arise 
if $H^1(\CC, \CN_\CC)$ is nontrivial.
For simplicity, we only consider
a one-parameter deformation.  If one chooses a small parameter $\epsilon$, 
and tries to find a finite holomorphic
deformation order
by order in $\epsilon$, one computes that the
obstruction to finding a solution at each order
is represented by an element of $H^1(\CC, \CN_\CC)$ \kodaira.
In particular, if there is a nonzero obstruction at order $\epsilon^n$,
then this geometry is naturally described by a superpotential
of the form $W = \Phi^{n+1}$.\foot{Of course the correct
normalization of the fields, and thus of the superpotential,
depends on the K\"ahler metric, which we will not compute in
this paper.}

The simplest example of such obstructed curves
begins with a threefold with $n$ isolated curves,
as a particular 
deformation of the complex structure
causes these curves to coincide.
They become a single curve $\CC$ of multiplicity
$n$, and this curve has an obstruction at order
$n$ to holomorphic deformations.  
Physically this is described by $n$ massive vacua coalescing
into a single vacuum with superpotential $\Phi^{n+1}$.\foot{The situation
is actually a bit more complicated than this.  Our assertion only pertains
to the case $\CN_\CC=\CO\oplus\CO(-2)$.  It is an open problem
to classify the possible superpotentials that can yield a single curve with
multiplicity $n$, even in the next simplest case 
$\CN_\CC=\CO(1)\oplus\CO(-3)$.  An example in this case is the superpotential
$W(\Phi,\Psi)=\Phi^2\Psi+\Psi^3$, corresponding to a curve with 
multiplicity~4.}

The B-type examples we will study realize this construction
from the following starting point \refs{\twopI,\kmp}.  
Begin with a Calabi-Yau threefold which contains a rational
curve $\CC$ fibered over a genus-g curve $\CS_g$.
One may canonically associate an element
$\omega \in H^{1,0}(\CS_g)$ with a non-toric\foot{We will be
studying hypersurfaces in weighted projective space: for
these examples the non-toric complex structure deformations 
are those which are not monomial deformations of the defining
equation.} first order 
deformation of the complex structure.  The differential $\omega$ generically
has $(2g-2)$ simple zeros, which correspond to
the isolated rational curves in the family
$\CS_g$ which survive the deformation.  At points of positive codimension
in complex moduli,
these curves can coincide and form curves of higher multiplicity.\foot{In all
of these cases, $\CN_\CC=\CO\oplus\CO(-2)$, so our previous discussion 
applies.}
There is a natural superpotential for a single
D5-brane wrapped on some fiber $\CC_z$ over a point
$z\in \CS_g$, described
in \kklm.  The local modulus
$\phi$ of the rational curve $\CC_z$ over $z$ can be thought
of as an element of the holomorphic tangent space
$T_z^{1,0}\CS_g$, and it is the scalar
component of a superfield $\Phi$.  The superpotential
is then:
\eqn\superpot{
	W(\Phi;\omega) ~=~\langle \omega,\Phi\rangle 
	~+~{1\over 2!} \langle\partial \langle \omega, 
	\Phi\rangle,\Phi\rangle ~+~ 
	{1\over 3!}\langle \partial \langle \partial 
	\langle\omega,\Phi\rangle,\Phi\rangle,\Phi\rangle ~+~\cdots\ ,
}
evaluated at $z$.  Here $\langle ~,~\rangle$ is the usual inner product between
forms and vectors, and $\partial$ is the Dolbeault operator on $\CS_g$.
It is easy to see that the expansion in \superpot\ can be truncated
after $(2g-1)$ terms without changing the location and multiplicity of
the critical points of $W$.

Below are several examples which realize this
general framework.  These will be our testing ground
for a discussion of open-string mirror symmetry.
The considerations of \kklm\
yield some predictions for the mirror three-cycles
which we will give at the end of these examples:
we will describe and explore the mirror examples in \S3.

\bigskip
\noindent{\it{Ur-Example}}

Our examples will all be orbifolds of 
the Calabi-Yau hypersurface $W$ 
of degree 8 in $\IP^{4}_{1,1,2,2,2}\ $, defined 
for example by the equation
\eqn\eqexone{
	p = z_1^8 + 
	z_2^8 + z_3^4 + z_4^4 + z_5^4 = 0.
}
$W$ has a singularity at $z_1 = z_2 = 0$, 
inherited from the ambient weighted projective space.  
Blowing up this $Z_2$ singularity yields a 
family of $\IP^1$s, parametrized by the genus
3 curve $\CS_3$:
\eqn\curve{z_{3}^4 + z_{4}^4 + z_{5}^4 = 0.} 
The non-toric deformations associated to $H^{(1,0)}(\CS_3)$
generically lift the family $\CS_3$ of $\IP^1$s,
leaving four isolated $\IP^1$s.
 
One can see the non-toric deformations explicitly,
by considering an equivalent description 
of $W$ as a complete intersection 
of a quartic and a quadric in $\IC\IP^5$ following \twopI.
One sees the equivalence by setting the 
homogeneous coordinates $(y_0,\cdots,y_5)$ of
$\IC\IP^5$ equal to $(z_1^2,z_2^2,z_1 z_2,z_3,z_4,z_5)$.  
Then the quadric  equation
\eqn\quadric{y_2^2 = y_0 y_1} 
of rank three is automatically satisfied.  
The model in $\IP^5$ obviously has complex structure moduli which
deform \quadric\ to an equation of
higher rank.  If one deforms the quadric to 
have rank four or fewer, then one is still
describing points in the complex structure 
moduli space of $W$.\foot{To see this, note that both \quadric\
and the rank~4 quadric $y_0y_1=y_2y_3$ can be desingularized by the
same blowup $y_0=y_2=0$, hence both blowups fit into the same moduli space.}
Deformations to quadrics of rank greater than four 
correspond to making an extremal transition from 
$W$ to another Calabi-Yau space. 
One finds a three-dimensional space of 
deformations of the quadric which leave one in
the moduli space of $W$, hence there are 
three non-toric deformations.  
This story is described in full generality 
(in the case of a family of $\IP^1$s 
parameterized by a genus $g$ curve, and the 
corresponding $g$ non-toric deformations) in \kmp. 

It is evident that when one deforms \quadric\ 
by a term which increases the rank to
four, e.g. $y_3^2$, one destroys the family of 
$\IP^1$s.  This is because the family is
located at $y_0  = y_1 = y_2 = 0$ (which is the same as
$z_1 = z_2 = 0$); the addition of $y_3^2$ to \quadric\ would then
force $z_3 = 0$ also.  But then the former genus 3 curve
\curve\ collapses to the four points $z_4^4 + z_5^4 = 0$.
Hence, instead of a one-parameter family there are 
now 4 isolated $\IP^1$s.

Upon wrapping a single D5-brane on a member of the family of
$\IP^1$s one finds a $U(1)$ gauge theory in four dimensions
with a single neutral chiral multiplet $\phi$ parameterizing
a local neighborhood in $\CS_3$.  After a generic
non-toric deformation described by $\omega$,
\superpot\ will describe a superpotential with four
massive vacua.

In the mirror manifold, $M$, the non-toric deformations
have the following description.  Toric K\"ahler
moduli in weighted
projective space arise from the volume of the
space, the blow-up parameters for the fixed loci of the
Greene-Plesser orbifold group,  and blow-up parameters for
any singularities of the weighted projective space which intersect
the Calabi-Yau.  If 
the CY hypersurface intersects one
of these loci $n+1$ times, the toric K\"ahler deformation changes the
size of all $n+1$ divisors simultaneously, while the remaining $n$
``non-toric'' moduli change the relative sizes.  

In our examples,
mirror symmetry demands the following statement
about the three-cycles $\Sigma\subset M$ mirror to 
$\CC\subset W$.  At the
locus in K\"ahler moduli space
with the three non-toric moduli turned off,
we have a unique first-order deformation which
by \refs{\mclean,\kklm} requires $b_1(\Sigma)\geq 1$.
(If the inequality is not saturated, the instanton
sum must give $b_1 - 1$ chiral multiplets a mass.)
The instanton-generated superpotential for one
of these moduli must vanish until the non-toric deformations
are turned on, and then the moduli space generically splits into
four massive vacua.

\bigskip
\noindent{\it{Example I}}

The first example which we will discuss is 
$W_1$, the orbifold of $W$ and $\CS_3\subset W$ by 
the discrete $Z_4\times Z_2\times Z_2$ group with generators
\eqn\firstgens{(1,i,i,i,i),~~(1,1,-1,-1,1),~~(1,1,1,-1,-1)\ .}
The family of $\IP^1$s on $W$ located at $z_1=z_2=0$ is orbifolded
by \firstgens, and the curve \curve\ becomes a genus 0 curve after orbifolding.
Because $\IP^1$ has no holomorphic one forms, 
this model does not admit non-toric deformations
which destroy the family of holomorphic spheres.  
Therefore, there is never a nontrivial
superpotential, for any complex structure.

Mirror symmetry requires that $M_3$
has a one-parameter family of
supersymmetric three-cycles.  This
would be most simply realized by a
family of three-cycles $\Sigma$ with $b_1(\Sigma) = 1$. 
No K\"ahler deformation of $M_3$ should
lead to a nontrivial disc instanton generated 
superpotential, so the family of three
cycles survives quantum corrections even after 
deformations of closed string K\"ahler moduli.

If $M_1$ is the mirror of $W_1$, we will see in Appendix A that
the mirror cycles $\Sigma\subset M_1$ to our family of $\IP^1$s have
$b_1(\Sigma) > 1$.  Therefore we expect disc instanton effects to
give masses to $b_1(\Sigma) - 1$ of the moduli of $\Sigma$
predicted by the classical geometry.

\bigskip
\noindent{\it{Example II}}

Our next example, $W_2$, arises from orbifolding
$W$ and $\CS_3\subset W$ by
the $Z_4 \times Z_4$ discrete group with generators
\eqn\gens{(1,i,i,1,-1),~~(1,1,1,i,-i)\ .}

Once again we have a family of $\IP^1$s at $z_1 = z_2 = 0$.  The curve
\curve\ becomes a genus-1 curve $\CS_1$, after orbifolding by \gens.
Thus, if we
wrap a D5-brane around a member of this
family, 
there is a single parameter in 
the superpotential
$W$ 
associated to the holomorphic differential on the curve $\CS_1$.
The corresponding superpotential \superpot\ is just
\eqn\superher{W(\Phi) = c \Phi\ ,}
where $c$ is related to the magnitude of the non-toric blowup.  
When $c\neq 0$ there are
no supersymmetric vacua: the auxiliary field $F$ in 
the chiral multiplet is non-vanishing, and
since we are coupled to closed string theory the 4d 
gravitino gains a mass. 
This is in keeping with the fact that after the deformation, 
the holomorphic spheres have all disappeared.

In the absence of coupling to gravity, one can redefine 
the supercharges so that
the superpotential \superher\ does not break 
supersymmetry; it simply
adds a harmless constant to the Lagrangian.
This is reflected clearly in the geometry of 
the example.  In a local neighborhood of
the $g=1$ curve of $\IP^1$s in $W_2$, 
the manifold looks like a product of an  
$A_1$ ALE space and a $T^2$.  This local geometry
is hyperk\"ahler and so has a family of
complex structures, parametrized by an $S^2$.
Upon performing the non-toric deformation \superher, 
one can choose a different complex
structure so that there are $\it still$ 
holomorphic curves.  Since $W_2$ is {\it not}
hyperk\"ahler, this is prevented by the global
geometry at finite volume.  Hence,
global features of $W_2$ are important in 
determining that supersymmetry is broken,
a fact which clearly reflects the need to 
couple the D-brane worldvolume theory to gravity in order to diagnose 
the supersymmetry breaking.  

If $M_2$ is the mirror of $W_2$,
the mirror cycles $\Sigma\subset M_2$ to
our family of $\IP^1$s should also live
in a one-dimensional family, so that
$b_1(\Sigma)\geq 1$.
The non-toric K\"ahler deformation of
$M_2$ breaks supersymmetry entirely via
disc instanton effects.

\bigskip
\noindent{\it{Example III}}

Example III works much like Example I.
Let $M_3$ be the orbifold of $W$ by the $Z_2$
symmetry generated by $\tilde g$:
\eqn\vdef{\tilde g ~=~(1,1,1,-1,-1)}
We will consider B-type branes 
on the mirror  $W_3$ of $M_3$. In $W_3$ there 
is still a one-parameter family of $\IP^1$s, parametrized
by a $\IP^1$ (roughly obtained by orbifolding 
the genus 3 curve in $W$).  Again,
there is no superpotential for
this modulus for any value of the complex structure.

For the mirror A-cycle we will find below that 
$b_1(\Sigma) = 1$, so classical
results apply for all values of the
K\"ahler moduli of $M_3$.

\subsec{Qualitative features of superpotentials for A-type branes}

\noindent{\it Coordinates on the moduli space of
A-type branes}

Let $M$ be a general Calabi-Yau threefold, and 
$\Sigma\subset M$ a special Lagrangian three-cycle.
For simplicity, assume
$b_1(\Sigma) = 1$, and assume
there is a single holomorphic disc instanton $D$ bounded
by a representative $\gamma$ of the generating
class in $H_1(\Sigma)$.  The cycle
$\Sigma$ moves in a one-dimensional family in $\sigma$-model
perturbation theory; as discussed in \refs{\bundlemir,\kklm},
we can parameterize
this family locally by a modulus field $\phi$:
\eqn\phidef{\phi ~=~{\rm Area}(D) + ia}
where the area is measured in string units, 
and $a$ is an axion (the Wilson line of
the brane $U(1)$ gauge field around $\gamma$).

This was the picture given in \kklm, but a moment's
thought indicates that it should be modified.
Consider for example a special Lagrangian torus
in $T^6$.  There are clearly no holomorphic discs
bounding the cycles of $T^6$, but this
does not mean that there is no moduli space
for the special Lagrangian subcycle. 
Indeed there is a simple ansatz which
naturally generalizes the above.  Begin with
a reference three-cycle $\Sigma_0$ in some family $\Sigma_t$,
defined by an embedding $f_t:\Sigma_0 \into M$.
Choose a family of deformations constructed from a harmonic
form in some class in $H^1(\Sigma_0)$ \mclean.
Then choose some one-cycle $\gamma_0\in \Sigma_0$ whose class
in $H_1(\Sigma_0)$ is dual to this cohomology class via the metric.
As $t$ varies in the chosen family of deformations,
$f_t(\gamma)$ will sweep out some tube $T$ in $M$.
A natural coordinate $\phi$ is:
\eqn\tubemod{
	\phi = \int_T \omega + i \int_{\gamma_t} A_t
}
where $\omega$ is the K\"ahler form on $M$ and $A_t$
is a flat connection on $\Sigma_t$.  When the tube is
holomorphic, the real part is simply the area of the tube.

\bigskip
\noindent{\it Finding nontrivial superpotentials}

Before launching into a detailed discussion of 
specific three-cycles mirror to the above examples,
we would like to gain some general and
intuitive understanding of the form of the superpotentials
directly in the language of the three-cycle geometry.

Following \ov, the sum over multiple covers of $D$ yields a superpotential
\eqn\superone{W ~=~\pm\sum_{n=1}^{\infty} {e^{-n\phi}\over n^2}\ .}
The sign here depends on details of the fermion determinants
around the instanton solution \wittcs.
It follows from \superone\ that
\eqn\deriv{{\partial W \over \partial \phi} ~\sim~\pm \sum_{n=1}^{\infty} 
	{e^{-n\phi}\over n} = \mp{\rm log}(1-e^{-\phi}).
}
This has a single critical point at $\phi = \infty$,
which from \phidef\ is the open string analogue of
large radius.  Of course one expects that to
reach $\phi = \infty$, $M$ must be at some infinite-radius
point.

Now, suppose instead that we have $k$ disc 
instantons $D_i$ bounding the same homology class
$\gamma\in H_1(\Sigma)$.  $D_i$ may 
differ by homology classes in $M$.  Choose $\Re (\phi)$ to
be the area of $D_1$.  The exponential of the action
for instanton $D_i$ is that for $D_1$ times
a factor $q_i$, 
the exponential of the (complexified) volume of
$[D_i - D_1] \in H_2(M)$.  Finally, assume each
of these discs are isolated.
The resulting superpotential is:
\eqn\wcomp{W~=~\sum_{i=1}^{k} 
\sum_{n=1}^{\infty} \sigma_i {e^{-n\phi}\over n^2}q_{i}^{n}\ ,
}
where $\sigma_i$ is the sign of the $\sigma$-model
fermion determinant for the instanton $D_i$.
The supersymmetric vacua satisfy:
\eqn\wcompd{\sum_{i=1}^{k} \sigma_i 
	{\rm log}(1-q_i e^{-\phi}) ~=~ 0\ .}
This is equivalent to a polynomial equation in $e^{-\phi}$, 
whose degree is the greater of $\#\{i\mid\sigma_i=1\}$ and
$\#\{i\mid\sigma_i=-1\}$ .  Note that the critical points
need not all be at large radius.  
The precise locations of the critical points
depend on the closed string K\"ahler moduli through the $q_i$.  

We get similar results when we include
new disc instantons in the class $d\gamma$ for
varying $d$.  The general result is that 
if we have $k_i$ disc instantons in classes
$d_i \gamma$, then (assuming for simplicity that all the fermion determinants
are positive) there are $\sum k_i d_i$ 
supersymmetric vacua.  For families of discs
some open-string version of the Gromov-Witten
invariants of closed string instantons should
replace $k_i$.

The main point of this discussion is that
it is not difficult to imagine
one-parameter families of special Lagrangian 
manifolds which yield, after disc instanton 
corrections, a discrete set of supersymmetric vacua.
This is fortunate as the mirrors of the 
B-brane configurations
discussed in \S2.2\ must exhibit this behaviour.

Another lesson is
that string instanton effects alter our expectations of
the topology of our three-cycles.  The natural physical
measure of $b_1(\Sigma)$ within 
$\sigma$-model perturbation theory is
the number
of massless chiral multiplets for a single D6-brane wrapped
on this cycle.  However, disc instanton effects may well 
give some of these chiral multiplets a mass, in which 
case there is no obvious 
physical 
distinction between the original three-cycle and
a cycle with smaller $b_1$.

\bigskip
\noindent{\it{Special features of A-cycles arising
as real slices}}

We will be focusing on special Lagrangian 
three-cycles constructed as the fixed point
locus of antiholomorphic involutions acting on $M$,
in other words antiholomorphic maps 
$\sigma:z \to \bar{z}$ which square to the identity.
The standard example, which we will use
in every case, is the real slice
arising as the fixed point set of
$z_i \to \bar{z}_i$.  
In this case, for each holomorphic disc 
we get a conjugate 
holomorphic disc.  If 
$f: D \to M$ is a holomorphic map, we can
define its conjugate holomorphic disc 
$g: D \to M$ by $g(z) = \overline{f(\bar z)}$.
Upon gluing these discs together we find that
for any special Lagrangian submanifolds obtained
as fixed points of antiholomorphic involutions, holomorphic
disc instantons always come as two halves of a rational
curve in $M$.  Furthermore, it is natural 
to conjecture that as
this special Lagrangian cycle moves through
a family $\Sigma_t$, one may find a set of one-cycles 
$\gamma_t\in\Sigma_t$ whose images in $M$ sweep out
this rational curve.

The superpotential can be derived from a variant of
\wcomp. Let $\IP$ be the rational curve in question, and
$t$ be the integral of the complexified
K\"ahler form of $M$ over $\IP$.  Let $z$ be
the action of the instanton described by $f$;
the action for the instanton described by $g$
is then $t - z$.  Assuming the fluctuation determinants have the
same sign, the superpotential one gets from summing over multiple
covers is up to overall sign:
\eqn\superchop{
	W = {\rm Li}_2 (1 - e^{-z}) + 
	{\rm Li}_2 (1 - e^{-t+z})\ .
}
It is easy to see that this has a supersymmetric
vacuum at $z = t/2$.  At this point in the open-
and closed string moduli space the 
superpotential is that of the local model
in \ov.

\newsec{Constructing the mirror three-cycles}

The next step to fleshing out the mirror map 
for open strings is, of course,
to characterize the mirror map for the
submanifolds on which they end.  In this
section we will find explicit special Lagrangian
three-cycles mirror to elements of the
families of $\IP^1$s described in the examples above.

\subsec{Strategy for identifying mirror cycles}

At an arbitrary point in the closed string moduli space,
it will be fairly difficult to find explicit mirror cycles.
Instead we focus on loci of the moduli space with physical
and mathematical significance.  In our 
B-cycle examples, the family of $\IP^1$s around which 
we intend to wrap D5-branes are known to have zero volume at some
submanifold in the full (complex and K\"ahler) moduli space;
these points occur when the resolutions of the orbifold singularities
discussed above have been turned off (along with the
associated NS-NS 2-form moduli).
We may identify these points physically by studying BPS
D2-branes wrapped on the same cycles in type 
IIA string theory.\foot{All of our statements are at string
tree level so we can be cavalier about changing brane dimension
like this.}  These D2-branes form massless (vector)
multiplets at this discriminant locus
as guaranteed by the BPS formula.

In type IIB on the mirror CY, the BPS formula implies
that a wrapped D3-brane must become massless
at the mirror discriminant locus.  Since
the mass receives no closed string worldsheet instanton
corrections, we need simply
find the mirror discriminant locus and the
vanishing three-cycle via classical geometry.
We will discuss the identification of this pair
of cycles in $W \subset\IP^4_{1,1,2,2,2}$ and its mirror; 
the same logic leads to a similar 
identification in all of our examples.

The mirror manifold $M$ of $W$ is easily 
constructed using the Greene-Plesser construction
\gp.  One quotients $W$ by a suitable maximal group of scaling symmetries, 
leaving only two complex structure deformations of
$M$.  These can be 
represented by the coefficients of the
monomials $z_1 \cdots z_5$  and $z_1^{4}z_{2}^4$
in the defining equation for $M$.

Let us work on the locus in moduli space 
where the defining equation is:
\eqn\ydef{(z_1^4 - z_2^4)^2 - 2\epsilon 
	z_1^4 z_2^4 + z_3^4 + z_4^4 + z_5^4~=~0.}
Here we have set the coefficient of $z_1 \cdots z_5$
to zero; this subspace 
of the complex structure moduli space of $Y$ intersects the discriminant 
locus at $\{\epsilon = 0 ~,~ \epsilon = -2 \}$.  These two points 
in moduli space can be seen to determine 
the same CY manifold by redefining $z_2$ by an eighth 
root of unity.
Now, we want to construct a supersymmetric three-cycle 
which is mirror to  a member of the family
of $\IP^1$s on $W$, discussed in \S2.2.  At least near large
complex structure, 
$\delta z_{1}\cdots z_{5}$ is the complex 
deformation of $M$
mirror to the size of the projective space $W\IP^{4}_{1,1,2,2,2}$.
Furthermore, we can identify
$\epsilon$ as mirror to the modulus controlling 
the size of the exceptional $\IP^1$ in $W$, as explained in \twopI.
Therefore, we are looking for a
three-cycle $\Sigma\subset M$ which collapses as $\epsilon \to 0$.  
This identification of 
mirror moduli holds in the other cases as well.
We will find a particular such 
three-cycle $\Sigma$ as 
a component of the fixed point locus of a real
involution acting on the A-model CY.  
Some set of fixed points of the 
Greene-Plesser quotient intersect the three-cycle,
and the details of the resolution of these singularities
will determine the topology of $\Sigma$.

In the following subsections 
we study the mirrors of the examples 
considered in \S2.   Mirror symmetry reverses 
the order of increasing complexity, so we will examine the three examples
in reverse order.

\subsec{Example III}

In this example the three-cycle topology is the simplest,
since the fewest blowups are required.  Recall
that $M_3$ is the orbifold of $W$ by
$\tilde{g} = (1,1,1,-1,-1)$.  
The only fixed points are at $z_4 = z_5 = 0$, so we introduce
a second $\IC^*$ action and a new coordinate 
$z_6$, where the second $\IC^*$ acts by
$(z_4,z_5,z_6) \to (\lambda z_4, \lambda z_5, \lambda^{-2}z_6)$.
The defining equation is modified to:
\eqn\modif{(z_1^4 - z_2^4)^2 - 
	2\epsilon z_1^4 z_2^4 + z_3^4 + z_{6}^2 
	(z_{4}^4 + z_{5}^4) = 0,} 
so that the manifold is preserved by this second $\IC^*$.
Now consider the real involution:
$$\sigma: (z_1,\cdots,z_6) \to (\overline z_1, \cdots,
	\overline z_6)\ .$$
We obtain a three-cycle $N$ as the fixed point locus of
$\sigma$ ($N$ has two components, which are basically two copies of
the desired three-cycle $\Sigma$).  We 
will see below that it vanishes as $\epsilon\to 0$.
For the rest of this subsection, we take 
$z_1,\cdots,z_6$ to be real (since we wish to work
on the fixed point locus of $\sigma$).  

We use the $\IC^*$ action of the 
$\IP^4$ to set $z_2 = 1$; we will see presently that we will
not need to leave this coordinate patch. Furthermore, we define:
\eqn\newvaris{x = z_1^4,~~Q = z_3^4 + 
	z_6^2 (z_4^4 + z_5^4),~~A = 2\epsilon + \epsilon^2\ .}
Solving \modif\ for $x$ in terms of the remaining variables we find:
\eqn\xsol{x = 1 + \epsilon \pm \sqrt{A - Q}.} 
We have two branches of solutions for $x$, 
which meet when $Q = A$.  The real slice
includes only the region $A \geq Q$, 
since otherwise $x = z_1^4$ would be imaginary.

Next, we blow up the orbifold singularity
induced by $\tilde g$ using the language 
of symplectic quotients.  We introduce a 
new K\"ahler parameter $r$ in the following
``D-term equation'' (\cf\ \edphases):
\eqn\symp{\vert z_4\vert^2 + 
	\vert z_5 \vert^2 - 2\vert z_6 \vert^2 = r} 
in the description of the
full complex manifold $M$:
along the real locus we can 
dispense with the absolute values.
Note that in the full CY we have to gauge away the
$U(1)$ under which $(z_4,z_5,z_6)$ have charges
$(1,1,-2)$.  Since the $z_i$ are real on $N$, the 
only gauge 
transformation which acts as an identification on $N$ is
$(z_4,z_5,z_6)\mapsto (-z_4,-z_5,z_6)$,
i.e.\ the orbifold by $\tilde{g}$.

Next, we can solve \symp\ for $z_4$:
\eqn\zfour{z_4 = \pm\sqrt{r - z_5^2 + 2 z_6^2}.}
This gives two branches for $z_4$ 
which are glued together along the hyperbola
$z_5^2 - 2 z_6^2 = r$.  
Reality of $z_4$ requires that $r \geq z_5^2 - 2z_6^2$.  These conditions
bound $z_3^2,z_5^2,z_6^2$ from above, allowing us to stay on the patch
where $z_2=1$.

Consider the regime $0< \epsilon<<1$, $r > \epsilon$. 
The region $Q \leq A$ intersects the region of 
real $z_4$  in a region with the topology of a solid cylinder, as
pictured below. 
\ifig\tube{$z_4$ is real between the walls; $x$ is real inside the 
tube.}{\epsfxsize2.0in\epsfbox{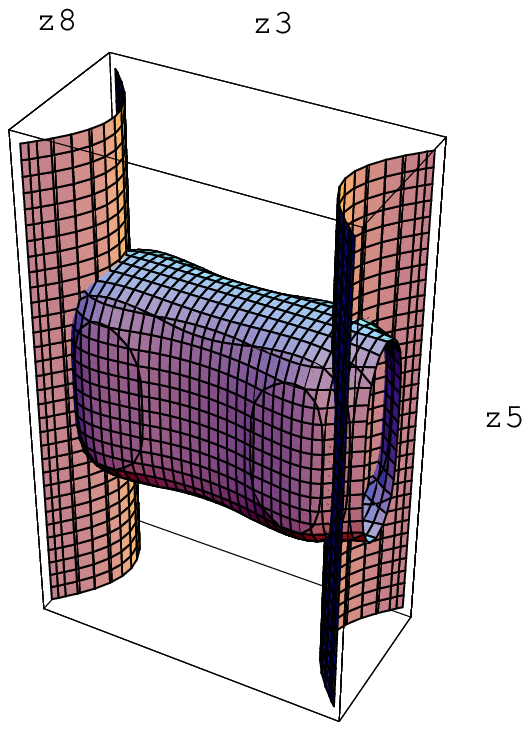}}
Making the orbifold identification merely 
halves the circumference of this cylinder.
The two branches of solutions for $z_4$ 
and the two branches of solutions for
$z_1^4$ give rise to four copies of this cylinder, 
which are glued along
the loci where the $z_4$ branches meet and the $z_1$ branches meet
(the sheet of hyperbolas and $Q=A$ respectively).   
The gluing along the boundaries of the $z_4$ branches yields two
solid tori; gluing along the boundaries of the $z_1$
branches then yields the closed three-manifold 
$\Sigma \sim S^2 \times S^1$, which has $b_1(\Sigma) = 1$.  
Note that $N$ consists of two copies of $\Sigma$, 
one with $z_1 > 0$, and the other with $z_1 < 0$ (we will abuse
notation and call both copies by the same name, since in any
case they are identical).

Because $\Sigma$ is a smooth special Lagrangian three-cycle with 
$b_1 = 1$, it is guaranteed by McLean's theorem to come in a family
of special Lagrangian cycles of real dimension one \mclean.  
Since $Q$ is positive semidefinite on the real slice, 
when $\epsilon\to 0$ 
the locus $Q \leq A$ collapses and the two components of 
$N$ are no longer finite-volume
three-manifolds.  We identify the two components of
$N$ with
two members of the family of $\IP^1$s in $W_3$. 

A D6 brane wrapping $\Sigma$ 
would naively yield, in the transverse 3+1 dimensions, a
4d ${\cal N}=1$ field theory with $U(1)$ 
gauge group and a single neutral chiral multiplet
$\phi$.  Although $\phi$ has no superpotential 
to all orders in sigma model
perturbation theory (this is the string theory analog 
of McLean's theorem), $\phi$
can receive a superpotential from disc instantons \kklm. 
In this case, we know from the mirror B-model 
geometry that there are no non-toric 
deformations which would lift the moduli space 
of supersymmetric $\IP^1$s.  This 
implies that there is no disc-generated superpotential in this case.

\bigskip
\noindent{\it{Disc Instantons}}

Some explicit examples of disc instantons with boundary 
on $\Sigma$ can be constructed in this example.
Consider the upper half plane parametrized by 
$u$.  Let $z_1,\cdots,z_6$ be given by
\eqn\ansatz{(z_1,\cdots,z_6) = (a_1 u, a_2 u,a_3 u^2,1,1,0)\ .}  
We take the $a_i$ to be real; this 
guarantees that the boundary of the disc (where
$u$ is real) is mapped to $\Sigma$.  

The disc must lie in $M_3$, which means that:
\eqn\acons{(a_1^4 - a_2^4)^2 - 2\epsilon 
	a_1^4 a_2^4 + a_3^4 = 0\ .}
Solutions to \acons\ provide holomorphic maps
into $M_3$ with boundary on $\Sigma$. 
In fact this ansatz yields a one-parameter 
family of discs: the constraint
\acons\ eliminates one of the $a_i$ 
and the freedom to rescale the $u$ plane
fixes another, but there is one free parameter left in the ansatz.
The fact that mirror symmetry implies that 
there is no disc-generated potential in this
case suggests that there is a cancellation between the contributions of 
different discs.  We will discuss such a mechanism in \S4.

\subsec{Example II}

The mirror $M_2$ of $W_2$ is constructed by orbifolding
$W$ by the $Z_4$ group generated by $g = (1,1,-1,i,i)$.  
The group element $g^2 = (1,1,1,-1,-1)$ is the symmetry by which we 
orbifolded in Example III, so we should still 
perform the resolution above.  However, $g$ 
itself fixes the locus $z_3 = z_4 = z_5 = 0$, 
which must be independently blown up.  
This is achieved by introducing another 
variable, $z_7$, and another $\IC^*$ action - 
the charges are summarized in the following 
table:

\halign{\qquad\qquad\indent#\qquad\hfil&\hfil#\quad\hfil&\hfil#\quad\hfil&
\hfil#\quad\hfil&\hfil#\quad\hfil&\hfil#\quad\hfil&\hfil#\quad
\hfil&\hfil#\quad\hfil\cr
&$z_1$&$z_2$&$z_3$&$z_4$&$z_5$&$z_6$&$z_7$\cr
$\IC^*_1$&$1$&$1$&$2$&$2$&$2$&$0$&$0$\cr
$\IC^*_2$&$0$&$0$&$0$&$1$&$1$&$-2$&$0$\cr
$\IC^*_3$&$0$&$0$&$2$&$1$&$1$&$0$&$-4$.\cr
}
\noindent
The defining equation is modified to 
\eqn\definm{(z_{1}^4 - z_{2}^4)^2 - 2\epsilon z_1^4 z_2^4 + 
z_7^2 z_3^4 + z_7 z_6^2 (z_4^4 + z_5^4) = 0.}

We will use $z_3$, $z_5$, and $z_6$ as coordinates 
on the real slice, $N$, 
which is the fixed point locus of 
$$\sigma: (z_1,\cdots,z_7) \to (\overline z_1, 
	\cdots,\overline z_7)\ .$$
Redefining
\eqn\newnewvar{x = z_1^4,~~Q = z_7^2 z_3^4 + 
	z_7 z_6^2 (z_4^4 + z_5^4),~~A = 2\epsilon + \epsilon^2}
we find: 
\eqn\xsolagain{x = z_2^4(1 + \epsilon \pm \sqrt{A - Q})\ .} 
On the real slice, the D-term equations for $\IC^*_{2,3}$ read
\eqn\Dterms{z_4^2 + z_5^2 - 2 z_6^2 - 
	r_2 = 0, ~~ 2 z_3^2 + z_4^2 + z_5^2 - 4 z_7^2 - r_3 = 0\ .}
Solving \Dterms\ for $z_4$ and $z_7$ we find:
$$z_4 = \pm\sqrt{r_2 - z_5^2 + 2 z_6^2}$$
and 
\eqn\wone{z_7 = \pm \sqrt{{1\over 2} 
	(r_2 - r_3 + z_3^2 + z_6^2)} \ .}
If we choose K\"ahler moduli so that $r_2 > r_3$, then
$z_7$ never vanishes, and the two branches of solutions 
never meet.  On the branch where 
$z_7 > 0$, $Q = z_7^2 z_3^4 + z_7 z_6^2 (z_4^4 + z_5^4)$ 
is positive semidefinite as in \S3.2, and 
\xsolagain\ tells us that 
this component of the real slice vanishes as 
$\epsilon \to 0$.  The other component of the 
real slice does not shrink on this locus and so is 
of no interest to us.  
Under the restriction that $z_4 = 
\pm\sqrt{r_2 - z_5^2 + 2 z_6^2}$ is real,
the three-cycle of interest again resides 
on the patch where $z_2 = 1$ (for this 
regime in closed string moduli space).  
The determination of topology 
goes through in complete analogy with Example III, and 
we again find two components of the real slice $N$, each of 
which is topologically $S^2 \times S^1$.  We again call
the components $\Sigma$. 

So we see again that a D6-brane on 
$\Sigma$ has a one-dimensional moduli space to all 
orders in $\sigma$-model perturbation theory.  
The non-toric deformation of $W_2$ which lifts 
the moduli space of supersymmetric $\IP^1$s must
in fact map to a small deformation of the K\"ahler 
structure of $M_2$.  This deformation cannot change
the topology of $\Sigma$, since $\Sigma$ is a smooth
three-cycle and the deformation can be made arbitrarily
small.  Hence, for the moduli spaces of the mirror pair
to match, the non-toric K\"ahler deformation must activate a 
disc-generated superpotential. 
We give further evidence for this below. 

\bigskip
\noindent{\it{Disc Instantons}}

For this example we can again construct explicit examples of disc 
instantons with boundary on $\Sigma$.
Using the holomorphic quotient description of $M_2$, 
we fix the three $\IC^*$ actions to set
$z_4 = z_5 = z_7 = 1$.  

Consider the upper half plane parametrized by 
$u$.  Let $z_1,\cdots,z_7$ be given by
\eqn\ansatzII{(z_1,\cdots,z_7) = (a_1 u, a_2 u,a_3 u^2,1,1,0,1)}
Again the $a_i$ are real, so that the boundary of the disc
$u\in \IR$ is mapped to $\Sigma$.  

In order that the disc lies in $M_2$, 
$a_{1,2,3}$ must again satisfy Eq. \acons.
As before,
we find a one-parameter family of holomorphic maps of the disc
into $M_2$ with boundary on $\Sigma$.

\subsec{Example I}

The mirror $M_1$ of $W_1$ is constructed by orbifolding
$W$ by the $Z_2\times Z_2$ group generated by $g_1 = (1, 1, -1, -1, 1)$
and $g_2 = (1, 1, 1, -1, -1)$.
This example provides the richest spectrum 
of phenomena for the A-cycles in $M_1$
as we have to perform the most blowups.  
Augmenting the weighted projective
space by the following additional variables and $\IC^*$
actions allows us to resolve 
all singularities which intersect the three-cycle:

\vfil\eject
\halign{\qquad\qquad
\indent#\qquad\hfil&\hfil#\quad\hfil&\hfil#\quad\hfil&
\hfil#\quad\hfil&\hfil#\quad\hfil&\hfil#\quad\hfil&\hfil#\quad\hfil&
\hfil#\quad\hfil
&\hfil#\quad\hfil\cr
&$z_1$&$z_2$&$z_3$&$z_4$&$z_5$&$z_6$&$z_7$&$z_8$\cr
$\IC^*_1$&$1$&$1$&$2$&$2$&$2$&$0$&$0$&$0$\cr
$\IC^*_2$&$0$&$0$&$0$&$1$&$1$&$-2$&$0$&$0$\cr
$\IC^*_3$&$0$&$0$&$1$&$0$&$1$&$0$&$-2$&$0$\cr
$\IC^*_4$&$0$&$0$&$1$&$1$&$0$&$0$&$0$&$-2$\ .\cr
}
\noindent
The defining equation for $M_1$ is:
\eqn\verymodif{
\eqalign{0 = (z_1^4 - z_2^4)^2 - 2\epsilon z_1^4 z_2^4 + 
z_7^2 z_8^2 z_3^4 + z_{6}^2 z_8^2 z_{4}^4 + 
z_6^2 z_7^2 z_{5}^4 \cr
\equiv (z_1^4 - z_2^4)^2 - 2\epsilon z_1^4 z_2^4 + Q\ .}} 
We will use $z_3$, $z_4$ and $z_5$ as our independent variables.  Solving 
the D-term equations associated to 
$\IC^*_2$, $\IC^*_3$ and $\IC^*_4$ for $z_6$, 
$z_7$ and $z_8$ on the real slice gives 
\eqn\twofour{ 
\eqalign{
	z_6 = \pm \sqrt{ {1\over 2} (-r_2 + z_4^2 + z_5^2)} \cr
	z_7 = \pm \sqrt{ {1\over 2} (-r_3 + z_3^2 + z_5^2)} \cr
	z_8 = \pm \sqrt{ {1\over 2} (-r_4 + z_3^2 + z_4^2)} .\
}}
where $r_{2,3,4}$ are the K\"ahler parameters controlling
the sizes of the associated
exceptional divisors in $M_1$.  


Choosing $r_{2,3,4} > 0$ we find a geometric phase.  
The real slice will have several identical components 
of which we choose one and call it $\Sigma$.
Since the function $Q$ defined above is positive 
semidefinite, this component will shrink when
$\epsilon \to 0$.

The analysis of the topology of this component $\Sigma$ of the 
real slice is included in the appendix.  The conclusion 
is that in the regime of K\"ahler moduli considered above, 
$b_1(\Sigma) = 5$.  In particular, $\Sigma$ is a connected sum of 5 
copies of $S^1 \times S^2$.  
The mirror $\IP^1$ has a one-dimensional moduli space,
while by McLean's theorem $\Sigma$ would move in a 5-dimensional
family.  Thus we are 
guaranteed the 
presence of a disc-instanton generated superpotential which 
lifts four of the flat directions.  

\newsec{Mirror symmetry and the superpotential}

We are interested in computing the superpotential
for the A-type examples in \S3, by finding
a mirror map for the open string moduli.  
In this section we will make some progress in this
direction.  We will start in \S4.1 by arguing that the
features of the mirror B-type examples near the
toric locus are captured by certain features of
disc instantons
in our A-type examples.  In \S4.2 we will
find a large-complex-structure limit of the
B-type examples for which the disc instantons
of the mirror three-cycle will be large,
and construct a mirror map for the
chiral multiplet in this limit.

\subsec{The superpotential near the toric locus}

Recall that for the Ur-example and for example II,
mirror symmetry requires the following story.
The ambient Calabi-Yau $M$ (resp.\ $M_2$) has $g$ non-toric deformations
with $g=3$ (resp.\ 1);
these arise because the hypersurface intersects $g$ of the
divisors of the (orbifolded) weighted projective
space twice, to create two divisors in the
hypersurface.  At the ``toric locus''
these divisors have the same size.  At this
point the three-cycles we study must have
a one-dimensional moduli space (namely a
genus-g curve).  As we leave this
locus, we acquire a superpotential and
are left with $2g-2$ isolated three-cycles.
We argue here that there will be different
disc contibutions which cancel on the toric locus.

In our A-cycle examples, the defining equation for the 
threefold $M$ may be written as:
\eqn\deform{ 0 = (z_1^4 - z_2^4)^2 - 2 \epsilon z_1^4 z_2^4 + Q\ ,}
where $Q$ is a function of all the variables $z_{k>2}$ 
other than $z_1$ and $z_2$.  It follows that on the hypersurface: 
\eqn\xone{ (z_1^4)_{\pm} = z_2^4 (1 + \epsilon \pm \sqrt{A - Q})}
with $A = 2 \epsilon + \epsilon^2$.  
Consider the map, $i: M \to M$, which fixes $z_{k>2}$
but flips the branches of $z_1^4$. 
We claim that this is an isometry of $M$ at the toric locus. 
Note that $i$ induces a map
on the toric part of the cohomology:
$i^*: H^2_{toric}(M) \to H^2_{toric}(M)$.
This happens to be the identity and so
preserves the K\"ahler class on $M$.
Furthermore, it preserves the complex structure, 
and so by Yau's theorem \yau\ it preserves the metric on $M$.  

The non-toric K\"ahler deformations 
are odd under $i^*$.  To see this, let us describe them
in more detail following \kmp.  In our examples
$M$ is a hypersurface in the quotient 
$\IP^4_{1,1,2,2,2}/\Gamma$, where $\Gamma$ is 
the relevant Greene-Plesser (GP) (sub)group.  
If $\Gamma$ has elements which fix the locus
$z_3 = z_4 = z_5 = 0$ (and not just 
varieties which contain it such as $z_3 = z_4 = 0$), 
this locus must be blown up to
desingularize $M$.  The K\"ahler
parameters controlling these blow-ups
have a natural mirror description in toric geometry
\refs{\twopI,\kmp}.
Within the lattice of exponents
of monomials in the ambient space of $W$,
the allowed monomials (preserved
by the subgroup of the
GP group complementary to $\Gamma$)
lie on a polyhedron.  The lattice points
on the lines and faces of this polyhedron correspond to
divisors in $M$ and monomial deformations of
$W$.  In particular the
lattice points corresponding to
$z_{3,4,5}^4$ in $W$
form the vertices of a triangular
face of this polyhedron and control the 
monomial deformations of the Riemann surface
$\CS_g \subset W$.  The $g$ interior
points can be used to construct the holomorphic
differentials of $\CS_g$, so they are
associated to the non-toric deformations
which destroy the family of $\IP^1$s over $\CS_g$.

In $M$ these $g$ interior points denote
the K\"ahler parameters controlling the blowup of
$z_3 = z_4 = z_5 = 0$.  The exceptional divisors
intersect the hypersurface twice in $M$,
at the loci $(z_1^4)_\pm = z_2^4 ( 1 + \epsilon \pm \sqrt{A} )$.
The non-toric moduli control the relative sizes of these
divisors in $M$.
Since the map $i$ defined above interchanges these two loci, 
and the non-toric deformations 
change their relative sizes, $i$ can no longer be 
an isometry away from the toric locus.  Furthermore,
at the toric locus $i$ will change the sign of the
non-toric deformation.

The map $i$ lifts naturally to a map on holomorphic discs. 
Hence, these discs 
should come in pairs related 
by $i$.
It is plausible that on the toric locus, the sign of the
contribution in Eq. \wcomp\ is changed by $i$;
then the disc instanton contributions of the
pair will cancel in the superpotential.
This could come about through the action  of $i$
on fermion zero modes or on the 
pfaffian.
When $i$ is not an isometry, 
the areas of the discs related by $i$ will differ, so
their contributions to $W$ will no longer cancel.

Note that this cancellation cannot happen for every disc
at the toric locus.  In particular, in example I
we find that $b_1(\Sigma) = 5$ while the true moduli space
is one dimensional.\foot{At special degenerate points in
the complex structure moduli space of $W$, the
dimension of the moduli space of the mirror $\IP^1$ enlarges
to two, but never to five.}
This suggests that the  
involution $i$ changes the sign of the contribution of discs
with boundaries in only one class of $H_1(\Sigma)$.
There is clearly much to understand here.

The set of examples we have considered fits into 
the more general framework discussed 
in \kmp.  The B-model CY in general contains a family of $A_N$ singularities 
fibered over a genus $g$ curve.  Upon resolving this family, one then 
finds $N$ families of $\IP^1$s.  
There are $g$ interior points 
on the relevant face of the toric diagram, and therefore $g$ non-toric complex 
structure deformations which destroy each family.  Furthermore, the defining 
equation for the A-model CY will be of degree $N+1$ in one of the 
variables, $x$, which is single-valued under the GP orbifold 
(the analog of $x=z_1^4$), leading 
to $N+1$ branches of solutions.  $g$ toric divisors each intersect these 
$N+1$ branches once.  On the toric locus, the Galois group, $S_{N+1}$, of the 
defining equation will act via isometries on the CY by interchanging the 
branches of solutions for $x$, leading to cancellations between discs.  
Upon turning on the non-toric K\"ahler 
moduli, these isometries are broken allowing a nontrivial superpotential.

\subsec{The mirror map}

One of our long-term goals is to use
mirror symmetry to find the explicit form of
the instanton sums for A-type branes.
Of course this sum is automatically computed
in the B-model, but we require a mirror map
for the open string fields in order that
this be of any use.
In the context of our models, this means the following.  
The B-brane moduli space is parametrized by
the complex coordinate $z$ on a genus $g$ surface, while 
the A-brane moduli space is 
parametrized via $\phidef$ by the disc area $A$ and
Wilson line $a$. 
Thus, defining
\eqn\qdef{q = e^{-\phi}\ ,}
we would like to find a map $z(q)$.  

As in the closed string case, this will be easiest
around ``large radius'' or
``large complex structure'' points, in particular when
$\Re (\phi)$ is large so that the instanton
action is small and classical geometry is a reasonable guide.
We therefore search for a map in a region of large
radius of our IIA models, and in the mirror large complex
structure limit of our IIB models.  In this limit
we can identify the mirror of the large-disc limit
of the A-cycle moduli space with a particular point
on the B-cycle moduli space.  Finally, since
the superpotential is explicitly computable
in the B-model side, we can use our
previous intuition about the A-type superpotentials
to guess at an explicit mirror map in this limit.

We will work exclusively with the Ur-Example in this section.

\bigskip
\noindent{\it The large-complex-structure limit of $W$}

On the manifold $M$, we are interested in the 
large radius limit.  
In particular, the sizes of discs ending on the real slice 
are determined by sizes of rational curves (as 
shown in \S2.3), so
we demand that the exceptional divisors which intersect
the three-cycle are large.  
The mirror locus in the complex structure 
moduli space of $W$ is 
specified by the defining equation
\eqn\largecom{p ~=~\alpha z_3^2 z_4 z_5 + 
	\beta z_3 z_4^2 z_5 + \gamma z_3 z_4 z_5^2 = 0\ }
with $\alpha,\beta,\gamma$ large and at fixed ratios
(recall that these monomials correspond to the
toric divisors in $M_1$).
\largecom\ can be rewritten as
\eqn\rewrite{z_3 z_4 z_5 ~(\alpha 
z_3 + \beta z_4 + \gamma z_5) ~=~0\ .}
This degeneration is similar to the
large complex structure limit discussed in
\syz.  It is described by four $\IP^3$s
at $z_3 = 0$, $z_4 = 0$, $z_5 = 0$ and
$\alpha z_3 +\beta z_4+\gamma z_5 = 0$ which 
intersect as shown in the figure.
\ifig\miro{$M$ in the 
large complex structure limit.  
The lines represent $\IP^3$s labeled 
by which coordinate vanishes on them.  `345' 
denotes the one where $\alpha 
z_3 + \beta z_4 + \gamma z_5 = 0$.  This picture 
also accurately portrays the degeneration of the curve $\CS_3$ 
where $z_1 = z_2 = 0$, in which case the lines 
represent $\IP^1$s.}{\epsfxsize2.0in\epsfbox{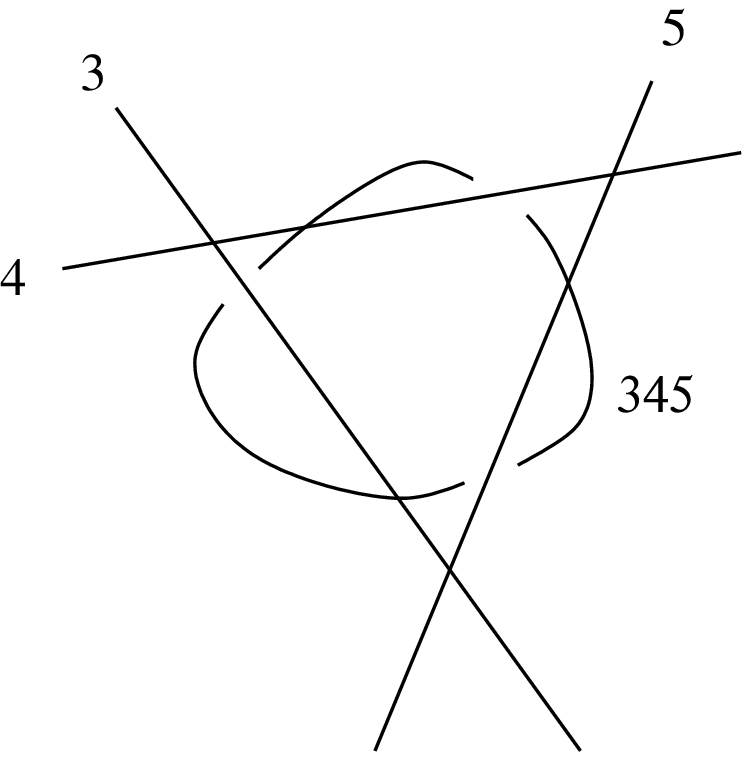}}
Each $\IP^3$ is identical to the others and
joins them in a symmetrical way.
As in \syz\ where the example of
the quintic at large complex structure is discussed, in 
this large complex structure limit of $W$ it is easy to write down a flat 
K\"ahler metric. Let us examine the metric on
$z_3 = 0$ near the locus $z_4 = 0$.  We may use the
$\IC^*$ action on the $W\IP^{4}$ to set $z_5 = 1$.
The standard residue formula yields the following expression
for the 
holomorphic $(3,0)$ form $\Omega$ in this degenerate limit:
\eqn\holoform{
	\Omega = \int_{p = 0}
	\frac{dz_1 \ldots dz_5}{p}
	= \frac{dz_1 dz_2 dz_4}{z_4 (\beta z_4 + \gamma).}
}
Near $z_4 = 0$ the Ricci-flat metric is clearly:
\eqn\rfmetriclim{
	dz_1^2 + dz_2^2 + \frac{1}{\gamma^2}d(\ln z_4)^2\ .
}
Elsewhere on the slice it is:
\eqn\rfmetric{
	dz_1^2 + dz_2^2 + d\zeta^2,
}
where
\eqn\bettercoord{
	\zeta = \frac{1}{\gamma} 
	\ln \frac{\beta z_4 + \gamma}{z_4}
}

$\CS_g$ in this limit lives at $z_1 = z_2 = 0$.  It is
a chain of four $\IP^1$s, one in each $\IP^3$ and
joined as in \miro.  The genus-3 structure is clear
from this figure.
The metric is simply induced from Eq. \bettercoord.
In particular it is clear that the different components
are joined along infinite cylinders, 
parameterized by $\ln z_i$ for $z_i \to 0$.
This then is the asymptotic moduli space
for a D5-brane wrapped on an element of
the family of $\IP^1$s at $z_3 = 0$ near $z_4 = 0$.

\bigskip
\noindent{\it The large-radius limit of $M$}

In the mirror $M$ there are three toric divisors
which are taken to be large: near the toric locus,
this means that all relevant divisors are large since 
the toric modulus controls the sum of the sizes of the 
exceptional divisors.

The moduli of the A-type branes are the areas of discs;
according to our discussion at the end of \S2.3\ these discs
live in pairs forming $\IP^1$s in the Poincar\'e dual class of
these toric divisors.  When the brane wraps a real
slice it bisects these $\IP^1$s.  Let us
parameterize the $\IP^1$ by an altitudinal angle $\theta$
and an azimuthal angle $\rho$.   The equator is $\theta = 0$
and the real slice intersects this
equator.  Let $2 \pi R$ be 
the circumference of the equator.  
$R$ will be
a function of $\alpha,\beta,\gamma$ via
the mirror map for closed strings.  
As
the three-cycle moves through its
moduli space, it will sweep out the 
$\IP^1$ by intersecting it at fixed $\theta$.
We choose the open-string
modulus so that $\phi$ is the area of the
smaller disc.  For $\theta$ close to zero,
a natural metric on the disc coordinate is:
\eqn\atypemetric{ds^2 = dA^2 + da^2 = 2\pi R^4 
	\cos\theta (d\theta)^2
	+ \frac{1}{(2\pi R)^2 \cos^2 \theta} (d\alpha)^2 \ ,
}
where $\alpha \in [0,2\pi]$.
Near $\theta = 0$ we can set $\cos\theta = 1$ and
this is clearly a cylinder, with the
periodic direction given by the Wilson line.

\bigskip
\noindent{\it{Asymptotic identification of the coordinates}}

Let us take $\beta,\gamma\in \IR$.
Up to overall normalizations of
the fields 
we roughly identify
\eqn\firstpass{
	\ln z_4 = \phi
}
due to the periodicity of the imaginary parts of
each side of this equation.  
This use of the periodicity is similar to the use
of monodromy properties in identifying the
closed string mirror map.  Note that the
correct normalization of the fields is
extracted from the so far unknown K\"ahler
metric. Thus our map is only good up to some 
overall constant.

\vfill\eject
\noindent{\it Computing the superpotential}

As reviewed in \S2, the deformation along the
non-toric locus can be specified by the
choice of a holomorphic differential.
Let $\tilde{p}$ be the defining polynomial for
$\CS_g \subset W$ at $z_1 = z_2 = 0$.  
The general holomorphic differential can be writtten as:
\eqn\gendiff{
	\omega = \int_{\tilde{p}=0}
	\frac{z_3 dz_4 dz_5 + z_4 dz_3 dz_5
		+ z_5 dz_3 dz_4}{\tilde{p}} 
	\left(a z_3 + b z_4 + c z_5 \right)\ .
}
On the locus $z_3 = 0$, this becomes:
\eqn\diffcomp{
	\omega = \frac{dz_5}{z_5(\beta z_4 + \gamma z_5)}
		(b z_4 + c z_5)
	+ \frac{dz_4}{z_4 (\beta z_4 + \gamma z_5)}
		(bz_4 + cz_5)\ .
}
We can concentrate on the region near $z_4 = 0$
by using the $\IC^*$ action of $W$ to fix $z_5 = 1$.
Then
\eqn\diffcomptwo{
	\omega = \frac{dz_4}{z_4 (\beta z_4 + \gamma)}
		(b z_4 + c)\ .
}
In the flat coordinates $x$ we write
$\omega = f(x)dx$.  As shown in \kklm,  
$W'(\Phi) = f(\Phi)$ for the
associated chiral multiplet.  This
superpotential clearly has a single
vacuum at $z_4 = -c/b$.  It is easy
to see that there is a single such vacuum
in each of the $\IP^1$ components of
$\CS_3$ in this limit, for a total of
$2g-2 = 4$ isolated vacua.

We wish to make contact with the large-disc limit
of the toric locus of the A-cycle moduli space.
Therefore we push the vacuum to infinite distance by
sending $c\to 0$.  In this limit,
\eqn\bsuper{W_\zeta (\zeta) =  b z(\zeta) = 
	\frac{\gamma}{e^{\gamma \zeta} - \beta}
}
for the superpotential of the B-brane.
If the A-cycle superpotential were dominated
by a single pair of discs, the corresponding superpotential
would be that in Eq. \superchop.  Certainly as
$z_4 \to 0$ and $\phi \to \infty$, \bsuper\ and
\superchop\ are equal to lowest order in $z_4$ and $e^{-\phi}$,
given \firstpass.

Our candidate superpotentials are equal only to lowest order as we
only have an asymptotic mirror map at present.  There are several
complications in constructing an exact mirror map.  First, given our
experience with Example I, we expect that the three-cycle has a
sizable number of classical moduli.  All but one gain masses from
instanton corrections, but the remaining moduli space may be a
nontrivial submanifold of the classical moduli space.  So our formulae
for the superpotential as a function of this modulus are undoubtedly
rather schematic.

Secondly, we have assumed that the only contributions
are a disc $D \in H_2(M,\Sigma)$ plus its three
images arising from the
anti-holomorphic involution and the map $i$
discussed in \S4.1.  Of course there may be higher-degree
contributions $nD$ which are not multiple covers,
and there may be contributions from discs $D'$ for
which $[D-D']$ lies in a nontrivial class in $H_2(M)$.
We leave these issues for future work.

Finally, we have not yet found a global topology of
the moduli space of A-branes 
which matches the topology of the moduli space of the
B-branes, even in this degenerate limit.

\newsec{Discussion}

In this paper, we have presented explicit
examples of three-cycles with nontrivial topology which
are mirror to two-cycles which have either obstructed holomorphic
deformations or no deformations.  After the results of \kklm,
this clearly shows that there is a disc instanton generated
superpotential for the moduli of such cycles.  
We have
certainly not given a complete formulation of the mirror map in these
examples, but we have made a first step
by presenting an analog of the monomial-divisor mirror
map for closed string moduli \mdmm.

Although we lack the full power of $\CN=2$
special geometry that exists for closed string mirror
symmetry, the structure of the $\CN=1$ theory
we are studying gives us some information.
In particular, the superpotential $W$ must
be holomorphic in the appropriate variables. 
Therefore, in limits where one has
an open string modulus $\phi$ which is periodic with
period $2\pi i$, the superpotential must be a
holomorphic function of $e^\phi$.
This together with some detailed knowledge of the
behavior of $W$ at singularities should be enough to determine the
function entirely. 

We can also draw some general lessons
from this work and the results of \kklm.
As with closed strings,
in making mathematical statements using mirror
symmetry one must take the instanton
corrections into account.  For instance,
there is a general conjecture that fundamentally, mirror
symmetry is a relation between 
the Lagrangian submanifolds of a threefold and 
the semistable coherent sheaves on its mirror
\konthom.  The fact that stringy nonperturbative
effects prevent generic special Lagrangian three-cycles
from being supersymmetric indicates that this
comparison will be complicated.

It would be interesting to study these issues in the presence of 
orientifolds (required for
tadpole cancellation), as a step
towards genuine model-building.\foot{We
would like to thank S. Sethi for a discussion
of these points.}  Many of the results of this paper
and \refs{\bdlr,\kklm} 
rest on the fact that from the $\sigma$-model point
of view, the superpotential is essentially a
topological quantity and can be computed in
an appropriately twisted theory.  Since
$\CN=2$ worldsheet supersymmetry is a consequence
of $\CN=1$, $d=4$ spacetime supersymmetry \bdfm,
the twisted theories will still make sense
in the presence of orientifolds.

Another subject worth exploring is the behavior of the
topology of a given 
special Lagrangian cycle $\Sigma$ as the closed string parameters vary.
It is clear from some of our examples that 
the topology of $\Sigma$ can change as one varies K\"ahler parameters 
of
the ambient Calabi-Yau space.  For instance, in the example we discuss
in Appendix A, different choices of the blow-up parameters $r_{2,3,4}$ 
yield three-cycles of different topology in the same homology class.

\bigskip
\centerline{\bf{Acknowledgements}}
We would like to thank M.R. Douglas, 
Y. Eliashberg, A. Klemm, D. Morrison, J.R. Myers, M.R. Plesser,
S. Sethi, C. Vafa, E. Zaslow, and Y. Zunger for helpful communications.
We thank P. Kaste for pointing out an error in an earlier version 
of this paper.
S. Kachru was supported in part by an A.P. Sloan Foundation Fellowship,
and by the DOE under contract DE-AC03-76SF00515.  S. Katz was supported
in part by NSA grant number MDA904-98-1-0009, and would like to thank
the Stanford High Energy Theory Group for their hospitality during the
course of this project.
A. Lawrence was supported in part by 
the DOE under contract DE-AC03-76SF00515
and in part by a DOE OJI grant awarded to E. Silverstein;
he would like to thank
the University of Chicago High Energy Theory Group,
the Rutgers High Energy Theory Group, and especially
the Oklahoma State University Department of Mathematics
for their hospitality during the course of this project.
J. McGreevy was supported in part by the Department of Defense
NDSEG Fellowship program.  
This project received additional support
from the American Institute of Mathematics.

\medskip

\appendix{A}{Some Details Concerning Example I}

In this appendix we determine the topology of $\Sigma$, the 
component of the real slice of the mirror of 
$\IP_{1,1,2,2,2}[8]$
on which all of the $w_i$ are positive.  
We will choose a regime in moduli space 
where 
\eqn\also{\epsilon >> r_i,  ~~~ r_1 >> r_i}
for all $i \geq 2$.

As in the simpler examples, we solve the defining 
equation for $x = z_1^4$ by
\eqn\xsolagainagain{x = 1 + \epsilon \pm \sqrt{A - Q}.} 
We have set $z_2 = 1$ again.  
We will see that the other variables charged under $\IC^*_1$ 
are bounded on $\Sigma$ and so $\Sigma$ is entirely 
contained in this coordinate patch.
The locus $B \equiv \{Q=A\}$ where the two branches of $x$ are joined 
is a big (not quite round) ball in the 
$\IR^3$ coordinatized by $z_{3,4,5}$.  
The branches of $x$ are then two copies of 
this ball glued along the boundary.  
The loci where the branches of $z_{6,7,8}$ 
are joined are three tubes surrounding 
the coordinate axes and ending on $B$.  
The region where all variables are real 
is the part of the inside of $B$ which 
is outside the union of these tubes.
Suppose we are in a regime of moduli where $r_3 + r_4 < r_2$.  
Then the tubes 
surrounding the $z_3$ and $z_4$ axes will both intersect the one 
surrounding the $z_5$ axis, but not each other, like this:  
\smallfig\figa{The real slice is the ball with the tubes removed.  
The tubes are labeled according to which branches are glued along 
them.}
{\epsfxsize2.0in\epsfbox{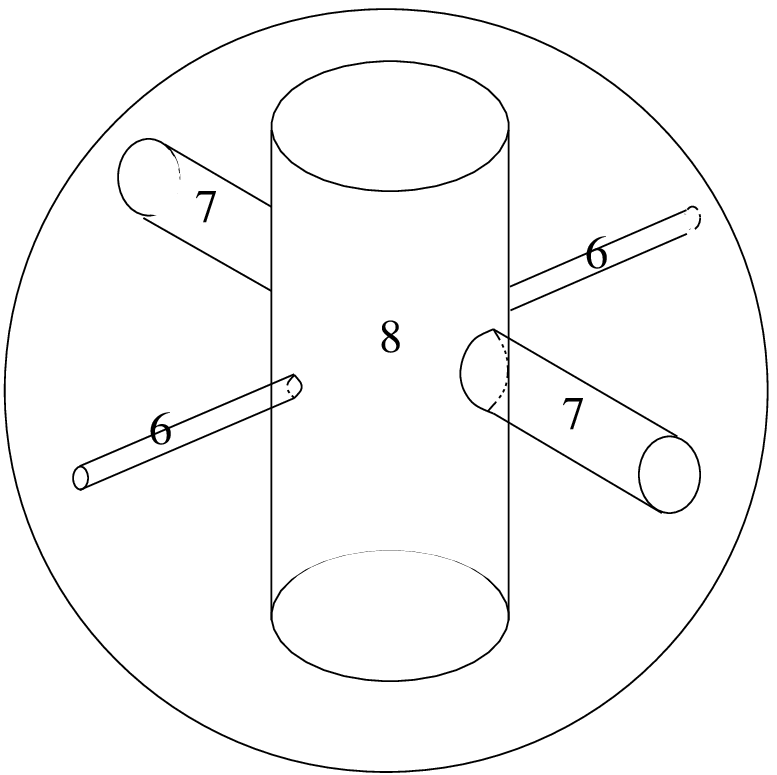}}

Now divide by the orbifold group which maps the real slice
to itself.  It acts by 
flipping signs in pairs:
$$ (z_3, z_4, z_5) \sim  (z_3, -z_4, -z_5)\sim  
	(-z_3, z_4, -z_5)\sim  (-z_3, -z_4, z_5).$$
\smallfig\figb{A fundamental domain for the orbifold action on the real 
slice (glue 
along the dotted lines with matching arrows).}
{\epsfxsize2.0in\epsfbox{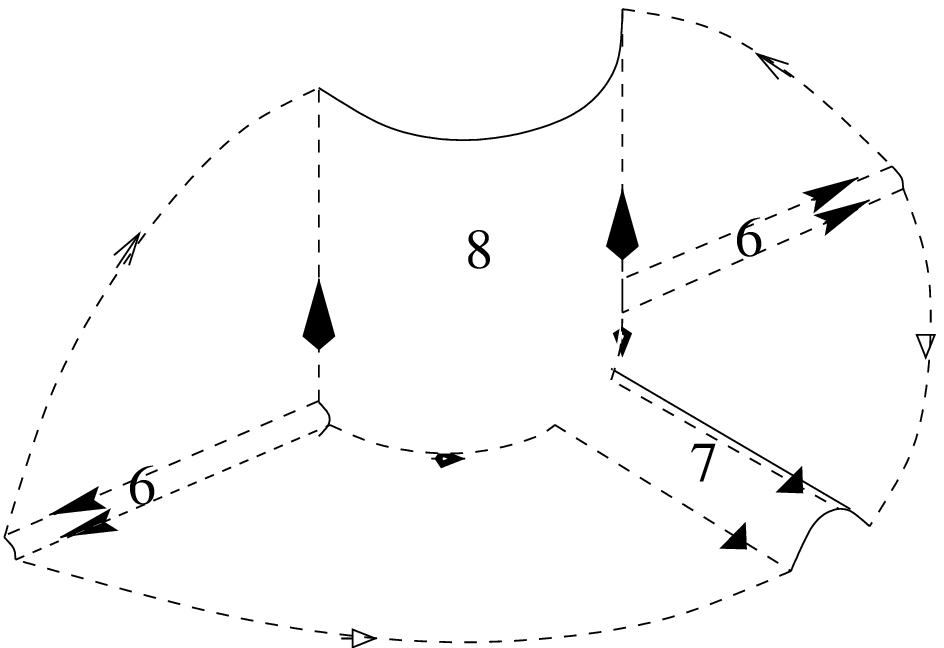}}
\noindent
After performing this identification, 
the plumbing fixture 
surrounding the origin depicted in \figa\
becomes a half-cigar (where $z_8=0$) 
ending on $B$
with two smaller tubes coming off of it (where $z_6=0$ and $z_7=0$ 
respectively) and ending on $B$ as well.

Next, glue the two $x$-branches along $B$.  This produces an $S^3$ 
with the following set removed:  The locus where $z_8$ becomes imaginary 
is now a full cigar, and 
the $z_6 = 0$ and $z_7=0$ loci are two 
handles coming off of this cigar.
\smallfig\figc{The real slice is the ball 
with the blob in the middle excised.}
{\epsfxsize2.0in\epsfbox{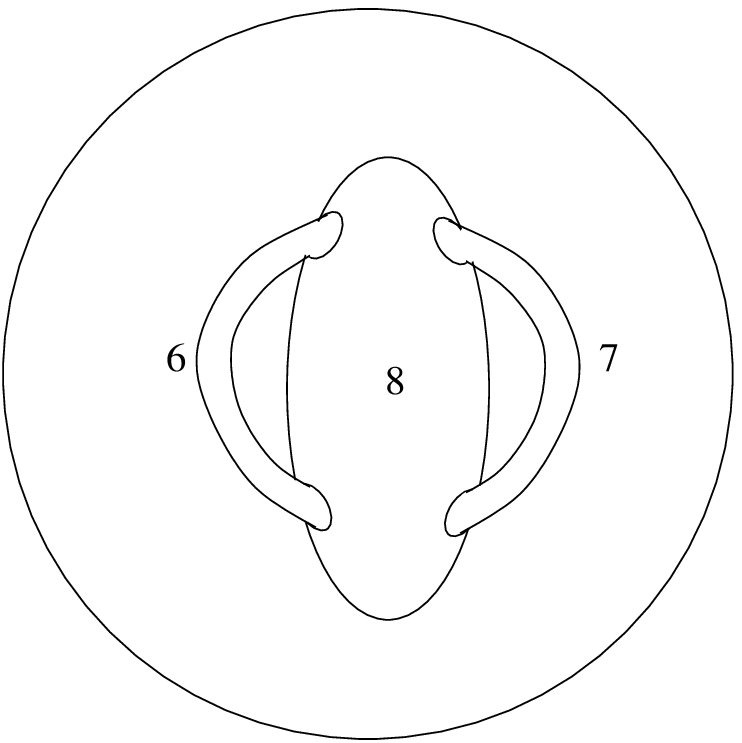}}
\noindent
An $S^3$ with the $z_8$-cigar removed is again a three-ball; the two handles 
coming off of the cigar become tunnels through this three-ball.  
\smallfig\figd{The previous picture turned inside-out.  The real slice is now 
the inside of the ball minus the two tunnels.  The boundary of the 
ball is where $z_8 = 0$.}{\epsfxsize2.0in\epsfbox{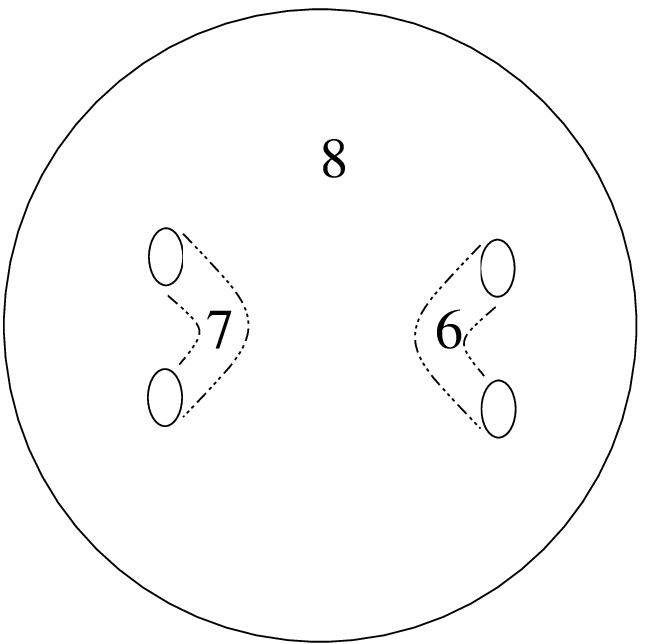}}

We take four copies 
of this creature to represent the two branches each of $z_6$ and $z_7$.  They are glued 
in pairs along the tunnels.  To see what this is we must 
use the fact that gluing handlebodies 
along a tunnel is the same as 
gluing along a tube that contains a handle.
\smallfig\fige{Gluing handlebodies (in particular, solid cylinders) along 
a tunnel is the same as connecting them via a tube with a handle in it.}
{\epsfxsize4.0in\epsfbox{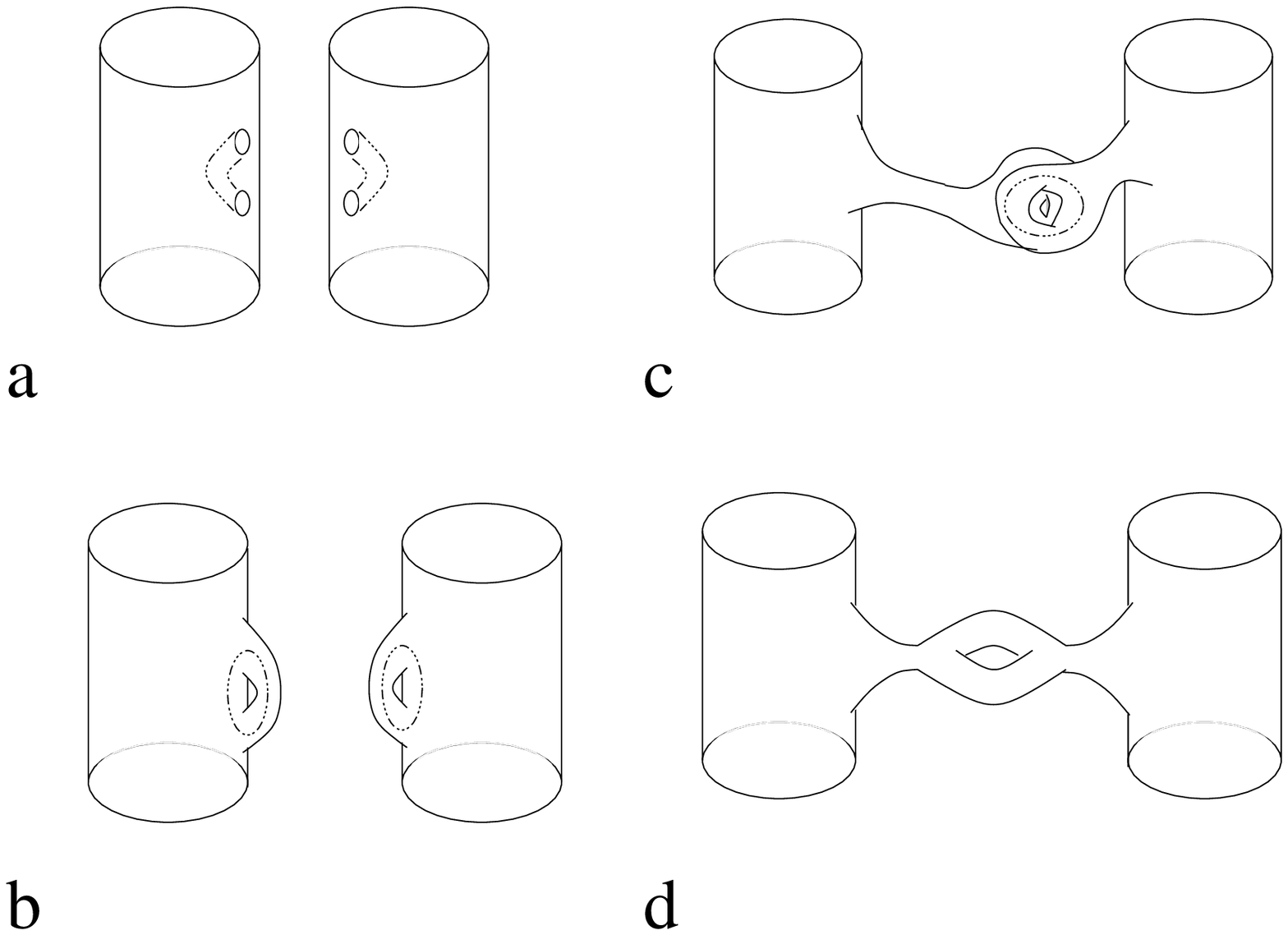}}

After we do this gluing, we find a solid genus $5$ surface for each branch 
of $z_8$.  
The boundary of this surface is where $z_8=0$.  
\smallfig\figf{Our special Lagrangian three-cycle is obtained by 
gluing two of these along their boundaries via the trivial 
identification.  The numbers 
along the top and left indicate which branch each ball represents.}
{\epsfxsize2.0in\epsfbox{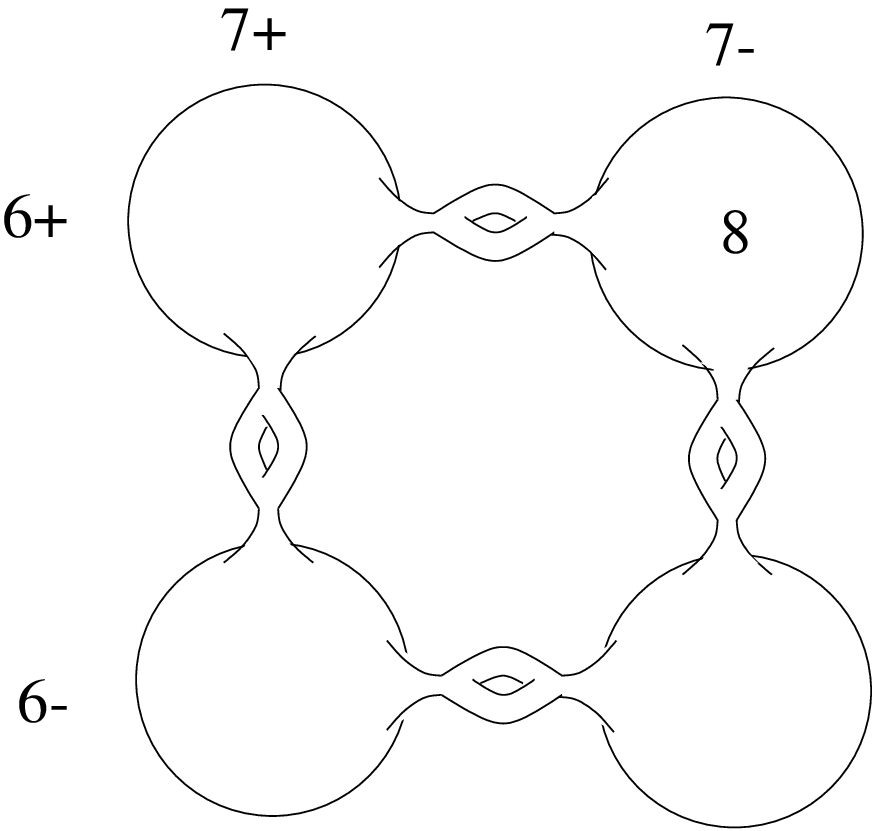}}

Since the two different branches for $z_8$ meet at $z_8=0$, we now glue
two copies of the genus $5$ surface together along their boundaries.
In general, two solid genus $g$ surfaces glued in this 
manner describe a Heegaard splitting of a connected sum of $g$ 
copies of $S^2 \times S^1$.  Hence, this three-cycle has $b_1(\Sigma)
=5$.

\listrefs 
\end